\documentclass{ws-ijmpe}
\usepackage[super,compress]{cite}
\usepackage{bm,color}
\begin{document}

\markboth{Yoritaka Iwata, Sophia Heinz}{Fissibility of compound nuclei}

\catchline{}{}{}{}{}

\title{Fissibility of compound nuclei}

\author{\footnotesize YORITAKA IWATA}

\address{GSI Helmholtzzentrum f\"ur Schwerionenforschung, Planckstrasse 1,  \\
Darmstadt D-64291, Germany \\
y.iwata@gsi.de}

\author{SOPHIA HEINZ}

\address{GSI Helmholtzzentrum f\"ur Schwerionenforschung, Planckstrasse 1,  \\
Darmstadt D-64291, Germany}

\maketitle

\begin{history}
\end{history}

\begin{abstract}
Collisions between $^{248}$Cm and $^{48}$Ca are systematically investigated by time-dependent density functional calculations with evaporation prescription.
Depending on the incident energy and impact parameter, fusion, deep-inelastic and fission events are expected to appear.
In this paper, a microscopic method of calculating the fissibility of compound nuclei is presented.
\end{abstract}

\keywords{Fissibility; Time-dependent density functional calculations.}

\ccode{PACS numbers:25.85.-w, 25.70.-z, 21.60.Jz}


\section{Introduction}
The synthesis of superheavy chemical elements~\cite{07hofmann,06oganessian} in the laboratory is one of the biggest challenges 
in nuclear physics.
It is an attempt for clarifying the existence limits of all the chemical elements, as well as the completion attempt of the 
periodic table of chemical elements.
We are concerned with heavy-ion collisions
\[  ^{248}{\rm Cm}~+~^{48}{\rm Ca} \]
with different impact parameters in this paper.
Let $A$ and $Z$ be the mass number and the proton number, respectively.
The neutron number $N$ is defined by $N = A -Z$, so that $N/Z$ of $^{248}$Cm and $^{48}$Ca are 1.58 and 1.40, respectively.
If fusion appears, $^{296}{\rm Lv}$ (= $^{296}116$) with $N/Z = 1.55$ is produced.

Fast charge equilibration~\cite{10iwata} is expected to appear in low-energy heavy-ion reactions with an incident energy of a 
few MeV per nucleon.
It provides a very strict limitation for the synthesis of superheavy elements.
Actually, the $N/Z$ of the final product is not above nor below the $N/Z$ of the projectile and the target (1.40 $\le$ N/Z $\le$ 1.58 in this case) in the case of charge equilibration, so that the proton-richness of the final product follows. 
Although the actual value of $N/Z$ depends on the two colliding ions, its value for the merged nucleus tends to be rather proton-rich for a given proton number of the merged system.
This feature is qualitatively understood by the discrepancy between the $\beta$-stability line and the $N=Z$-line for heavier cases. 
Superheavy compound nuclei are very fragile and fission is a very frequent channel which leads to disintegration of the compound 
nuclei even at low excitation energies.
In this paper, following the evaporation prescription shown in Ref.\cite{12iwata}, the fissibility 
(fission probability) of compound nuclei is calculated based on the time-dependent density functional theory (TDDFT).
The presented method brings about the microscopic derivation of the fissibility.
For reference, a possible fission dynamics based on the evaporation prescription was demonstrated in Ref.\cite{12iwata-2}.

\section{Methods} \label{sec1}
\subsection{Treatment of the thermal property}
Self-consistent time-dependent density functional calculations are employed in this paper.
TDDFT reproduces the quantum transportation due to the collective dynamics. 
In this sense, what is calculated by the TDDFT can be regarded as products after several $10^{-21}$~s, which corresponds to a typical time-scale of low-energy heavy-ion reactions (1000~fm/c), as well as to the inclusive time interval of any collective oscillations such as giant dipole resonance, giant quadrupole resonance and so on.
Meanwhile, thermal properties such as the thermal instability are not directly taken into account in TDDFT.
Indeed, the Skyrme type interaction used in TDDFT (for example, see Ref.~\cite{greiner-maruhn}) is determined only from several densities.
It is important that the most effective cooling effect arises from the emission of particles, and therefore it is expected that the break-up or fission of fragments including rather high internal excitation energy is suppressed in the TDDFT final products.
The additional thermal effects leading to the break-ups of fragments should be introduced.

Here is a fact that simplifies the treatment of thermal effects, that is, the difference of the time-scales.
Different from the typical time scale of low-energy heavy-ion reactions, the typical time-scale of the thermal effects is estimated by the typical time interval of collision-fission (fission appearing in heavy-ion collisions): several 10$^{-19}$~s.
It is reasonable to introduce an evaporation prescription simply to the TDDFT final products.
In this context the TDDFT final fragments have the meaning of products just after the early stage of heavy-ion reactions (several $10^{-21}$~s).

\subsection{Evaporation prescription}

In complete fusion reactions the cross-section for the formation of a certain evaporation residue is usually given by three factors \cite{anto}:
\begin{equation}
\sigma_{ER}(E_{cm}) = \sum_J \sigma_{CP} (E_{cm},J) \times P_{CN} (E_{cm},J) 
 \times P_{SV} (E_{cm},J)
\end{equation}
where $\sigma_{CP}$, $P_{CN}$ and $ P_{SV}$ mean the capture cross-section, the probability for the compound nucleus formation, and the probability for survival of the compound nucleus against fission.
All three factors are functions of the centre-of-mass energy $E_{cm}$ and the total angular momentum $J$, where $J$ can be related with the impact parameter. 
For light systems $P_{CN}$ and $P_{SV}$ are about unity and $\sigma_{ER}\,\approx\,\sum_J\sigma_{CP}$. 
But in superheavy systems the strong Coulomb repulsion and large angular momenta lead to small values of $P_{CN}$ and $P_{SV}$ and therefore to the small cross-sections of the evaporation residues observed in the experiments. 
This means, different from light systems, it is necessary to introduce additional thermal effects for the superheavy element synthesis.
First, $\sigma_{CP}$ is sufficiently considered in the TDDFT if we restrict ourselves to a sufficiently high energy exceeding the Coulomb barrier (cf. sub-barrier effects such as tunnelling are not taken into account in the TDDFT).
Second, $P_{CN}$ is fully considered in the TDDFT, which is a kind of mass equilibration also related to charge equilibration.
Third, $P_{SV}$ whose relative time-scale is by no means equal to the former two probabilities is not satisfactorily considered in the TDDFT.
This probability is much more related to thermal effects.
Consequently, further consideration is necessary only for $P_{SV}$ as far as the energy above the Coulomb barrier is concerned.

Several factors are included in $P_{SV}$ such as probabilities for fission of the compound nucleus,
neutron-evaporation, proton-evaporation, deuteron-evaporation, alpha-particle-evaporation and so on.
Probabilities of neutron and $\alpha$-particle emissions are considered by
\[ P_{SV} :=(1- P_{n, evap}) (1- P_{\alpha, evap}). \]
For the application of the evaporation prescription, see our preceding research summarized in Ref.~\cite{12iwata}.

\subsection{Treatment of the Fissibility}
First, a long-lived compound nucleus with a certain excitation energy is
formed.
The fissibility (fission probability) is obtained by additional TDDFT calculations (cf. the method shown in Ref.~\cite{12iwata-2}), 
where no initial velocity is given.
Here we are interested in the possibility of fission of a compound nucleus ($^{\mathcal A}{\mathcal Z}$): 
\begin{equation} \label{fiseq}
^{\mathcal A}{\mathcal Z}  ~\to~  ^{{\mathcal A}_1}{\mathcal Z}_1   ~+~  ^{{\mathcal A}_2}{\mathcal Z}_2,
\end{equation}
where $ {\mathcal A}_i$ and $ {\mathcal Z}_i$, which satisfy ${\mathcal A} = {\mathcal A}_1 + {\mathcal A}_2$ and 
${\mathcal Z} = {\mathcal Z}_1 + {\mathcal Z}_2$, are the mass numbers and the proton numbers of the fragments, respectively. 
First, a configuration of the two nuclei ($^{{\mathcal A}_1}{\mathcal Z}_1$ and $^{{\mathcal A}_2}{\mathcal Z}_2$) at a 
distance $R_0$ is defined as an initial state for additional calculations.
Second, choose $R_0$ systematically such that compound nuclei with several excitation energies are examined.
Note here that the TDDFT is a theory in which the total energy is strictly conserved, so that the total energy is conserved 
during the additional TDDFT calculations.
Third, the initial many-body wave function, which is given as a single Slater determinant, consists of single wave functions of 
two different initial nuclei, where a set of single wave functions are orthogonalized before starting TDDFT calculations 
(cf. the Gram-Schmidt orthogonalization method).

Fusion, fission and elastic scattering (never touched throughout the time evolution) can take place depending on the initial 
distance ($R_0$), because the excitation energy is associated with the initial distance.
In this way we obtain the possibility of fission as a function of the initial distance, and therefore as a function of the initial excitation energy.
This method brings about the microscopic derivation of the fissibility.
Note that the present method suggests the fissibility as purely due to the collective motion in which the nuclear and Coulomb forces are fully taken into account as well as the shell effects and the deformation.

\section{Fissibility of compound nuclei}  \label{sec2}

\begin{figure}[t] 
\centerline{\psfig{file=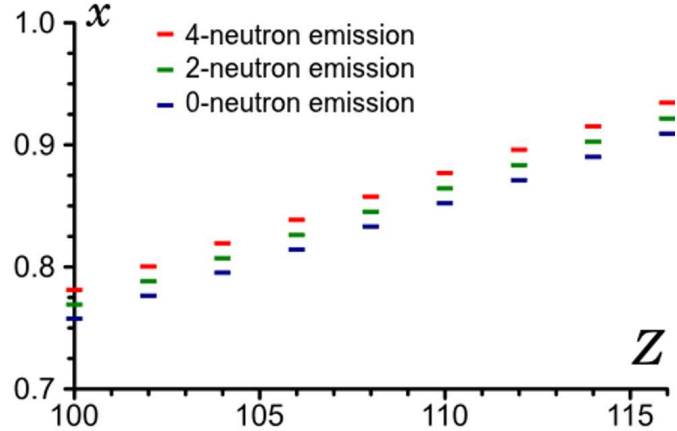,width=9.0cm}}
\caption{(Colour online) Fissility of compound nuclei based on the liquid drop model (Eq.~\eqref{fissility}).
Fissilities  of ground states are calculated for 0-neutron, 2-neutron, 4-neutron emission channels of $^{248}{\rm Cm}~+~^{48}{\rm Ca}$
and for the daughter isotopes of the respective decay chains.}
\end{figure}

\subsection{Fissibility based on the liquid drop model}
For comparison, let us introduce the phenomenological fissility~\cite{greiner-maruhn} ($0 < x \le 1$):
\begin{equation} \label{fissility}
 x = 
\left \{
\begin{array}{ll}
\displaystyle \frac{Z^2/A}{(Z^2/A)_C}; \qquad Z^2/A < 50,  \vspace{2.5mm}\\
1 ; \hspace{19.5mm} Z^2/A \ge 50,                 
\end{array} \right.
 \end{equation}
where the critical value $(Z^2/A)_C$ is roughly equal to 50.
In this equation the balance between the surface and Coulomb effects is taken into account within the liquid drop model description.
The value $x$ shows the stability against quadrupole deformation (at least); a nucleus can split up into two fragments if 
$x$ is close to 1.
Note that the fissility is a technical term generally used for the value of $x$, but it exactly means the fissibility.
Since this value expresses the instability against the quadrupole deformation, it is mostly related with the possibility of symmetric fission.
For $N/Z=1$ nuclei, $A=200$ is the threshold value whether $x < 1$ or not.
$x = 116^2/(296 \times 50) = 0.909$ in case of the Compound nucleus $^{296}{\rm Lv}$.
The fissilities of ground states for $^{296}{\rm Lv}$, $^{294}{\rm Lv}$, $^{292}{\rm Lv}$ and their several daughter nuclei (from different alpha-decay chains) are shown in Fig.~1 in which we found monotonous trends depending on both, neutron and alpha emissions. 
In this paper, instead of the fissility $x$, we present a method of calculating the fissibility in a microscopic way.

\begin{table}[pt] \label{table1}
\tbl{Values related with the possibility of fission.
Symmetric fission events are considered in the former two cases.
Asymmetric fission events related with the compound nucleus $^{296}{\rm Lv}$
(no neutron emission channel) are considered in the latter three cases.
The contact distance is calculated in the liquid drop model ($r_0 A^{1/3}$) taking into account the initial deformation of the TDDFT calculations.
The surface area of the resulting binary system, which is normalized by the value for $^{148}{\rm Ce}$ + $^{148}{\rm Ce}$, 
is also calculated in the liquid drop model.}
{\begin{tabular}{@{}lccc@{}} \toprule
Fission reaction & ${\mathcal Z}_1 \times {\mathcal Z}_2$ & Contact distance [fm] & Surface area (normalized) \\ 
\colrule
$^{296}{\rm Lv} ~\to~ ^{148}{\rm Ce}$ + $^{148}{\rm Ce}$ \hphantom{00} & \hphantom{0}3364& \hphantom{0}16.3 & \hphantom{0} 1  \\
$^{284}{\rm Ds} ~\to~ ^{142}{\rm Cs}$ + $^{142}{\rm Cs}$  \hphantom{00} & \hphantom{0}3025& \hphantom{0}16.1  & \hphantom{0} 0.97 \\
\colrule
$^{296}{\rm Lv} ~\to~ ^{40}{\rm Ca}$ + $^{256}{\rm Cm}$  \hphantom{00} & \hphantom{0}1920& \hphantom{0}13.7  & \hphantom{0} 0.93 \\
$^{296}{\rm Lv} ~\to~ ^{16}{\rm O}$ + $^{280}{\rm Hs} $  \hphantom{00} & \hphantom{0} 864& \hphantom{0}12.7  & \hphantom{0} 0.88 \\
$^{296}{\rm Lv} ~\to~ ^{4}{\rm He}$ + $^{292}{\rm Fl} $  \hphantom{00} & \hphantom{0} 228& \hphantom{0}11.5  & \hphantom{0} 0.83 \\
\toprule
\end{tabular}}
\end{table}

\subsection{Fissibility based on TDDFT calculations}
Before moving on to the calculation results, the two values related with the fissibility are shown for some cases treated in this paper (Table~1).
The value ${\mathcal Z}_1 \times {\mathcal Z}_2$ is related with the instability due to the Coulomb energy, while the 
contact distance is related with the threshold energy, depending on whether the nuclear force is operational or not.
In the case of Eq.~\eqref{fissility}, only the first value is taken into account by assuming the operation 
of the nuclear force.

Time-dependent density functional calculations with a Skyrme interaction (SLy6~\cite{Chabanat-Bonche}) are carried out in a spatial box of 48~$\times$~48~$\times$~24 fm$^3$ with periodic boundary condition.
The unit spatial spacing and the unit time spacing are fixed to 1.0~fm and 2/3~$\times$~10$^{-24}$s, respectively.
The initial states, which correspond to the ground states, are prepared by static density functional calculations.

In order to show a procedure of calculating the fissibility, we have several examples.
At first, the possibility of symmetric fission is investigated for compound nuclei included in an alpha-decay chain.
Here we consider the 0-neutron emission channel:
\begin{eqnarray} 
 ^{248}{\rm Cm}~+~^{48}{\rm Ca}  &&~\to~ ^{296}{\rm Lv} \nonumber \\ 
 && ~\to~ ^{292}{\rm Fl} + \alpha  
 ~\to~ ^{288}{\rm Cn} + 2 \alpha  
 ~\to~ ^{284}{\rm Ds} + 3 \alpha  
 ~\to~ \cdots  \nonumber
\end{eqnarray}
as a limiting case.
More concretely let us study the symmetric fissibilities of $^{296}{\rm Lv}$ and $^{284}{\rm Ds}$ corresponding to the cases with $^{{\mathcal A}_1}{\mathcal Z}_1 ~=~ ^{{\mathcal A}_2} {\mathcal Z}_2 ~=~ ^{148}{\rm Ce}$ and $^{{\mathcal A}_1}{\mathcal Z}_1 ~=~ ^{{\mathcal A}_2}{\mathcal Z}_2 ~=~ ^{142}{\rm Cs}$ (in Eq.~\eqref{fiseq}), respectively.
Next, the possibility of asymmetric fission is investigated for the compound nucleus $^{296}{\rm Lv}$.
Although the fissility $x$ is the value for symmetric fissions, the fissibility can be considered using the present method. 
The following fission events are considered:
\[ \begin{array}{ll} 
^{296}{\rm Lv} ~\to~ ^{256}{\rm Cm} ~+~ ^{40}{\rm Ca},   \\
^{296}{\rm Lv} ~\to~ ^{280}{\rm Hs} ~+~ ^{16}{\rm O},    \\
^{296}{\rm Lv} ~\to~ ^{292}{\rm Fl} ~+~ ^{4}{\rm He},    \\

\end{array} \]
where the mass asymmetries are $0.16$, $0.06$ and $0.01$, respectively. 

\begin{figure}[t] 
\centerline{\psfig{file=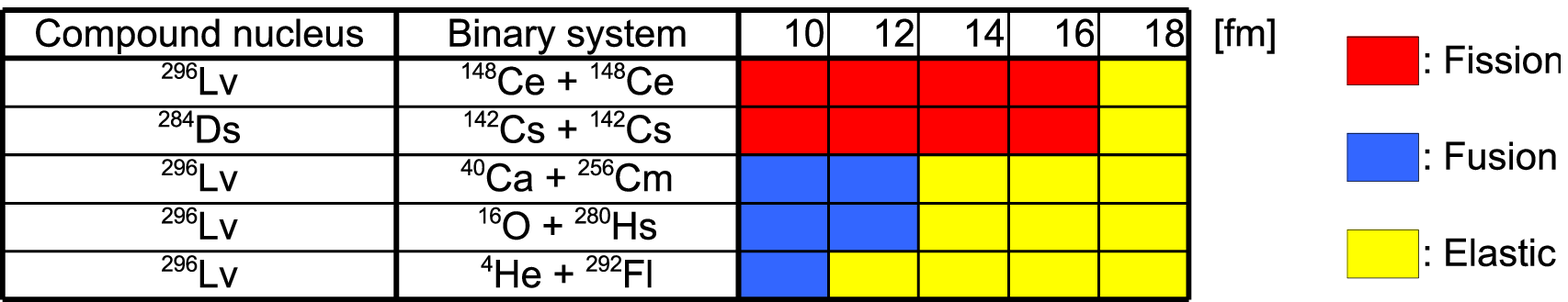,width=11.0cm}}
\caption{(Colour online) Systematic TDDFT results for two compound nuclei.
Results starting with different binary systems are shown for different initial distances $R_0 = 10, ~12, ~14, ~16, ~18$~fm. 
The initial compound nucleus is set to $^{296}{\rm Lv}$ except for a symmetric fission: 
$^{284}{\rm Ds} ~\to~ ^{142}{\rm Cs}$ + $^{142}{\rm Cs}$.}
\end{figure}

The systematic TDDFT results with different $R_0$ values are summarized in Fig.~2.
It is notable that no fission appears in all asymmetric cases, where the attraction due to the nuclear force is 
considerably larger than the repulsion due to the Coulomb force.
It means that the emission of small nuclei such as alpha-particles (alpha-decay) cannot be explained by collective dynamics in which the competition between the surface tension and the Coulomb force are fully taken into account.
In this context, another effect such as the thermal instability and the resulting cooling effect~\cite{12iwata} might play a role in the emission of small nuclei, instead.    
It is also suggested that there is a transition of fission mechanism of compound nuclei; from the mechanism mostly due to 
collective dynamics to another mechanism.
According to Table~1, the Coulomb energy of symmetric fission 
($ \min ({{\mathcal A}_1}/{{\mathcal A}_2},~{{\mathcal A}_2}/{{\mathcal A}_1}) = 1$) is 1.75 times larger than that of 
asymmetric fission with $ \min ({{\mathcal A}_1}/{{\mathcal A}_2},~{{\mathcal A}_2}/{{\mathcal A}_1}) =0.16$, while the surface energy ratio between them is only 1.08. 
It implies that the transition of fission mechanism arises from the different mass dependence between the surface and Coulomb effects.
This transition is identified by the threshold value for the mass asymmetry ratio.

\begin{figure}[t] 
\centerline{\psfig{file=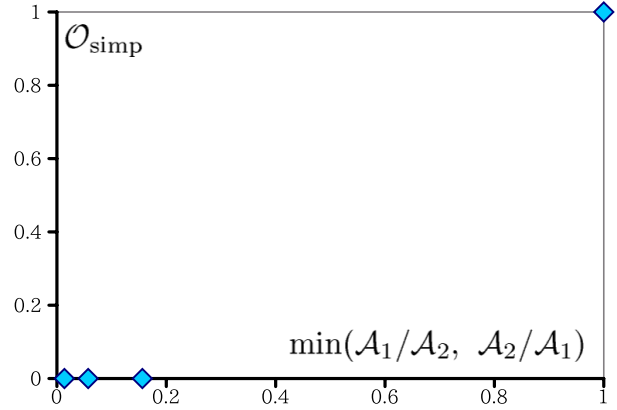,width=8.0cm}}
\caption{(Colour online) Fissibility of compound nuclei. 
${\mathcal O}_{\rm simp} = 0$ for very mass-asymmetric cases, while ${\mathcal O}_{\rm simp} = 1$ for mass-symmetric cases. 
There exists a jumping from ${\mathcal O}_{\rm simp} = 0$ to $1$ in the intermediate mass-asymmetry region: $0.16 < \min ({{\mathcal A}_1}/{{\mathcal A}_2},~{{\mathcal A}_2}/{{\mathcal A}_1}) < 1$.}
\end{figure}

For the results of symmetric fissions shown in Fig.~2, no significant difference is noticed between the two cases.
In those calculations the incrementation of $R_0$ is fixed to 2~fm, which is too large to identify the difference between the two cases.
On the other hand, difference in the results of asymmetric fissions shown in
Fig.~2 can be essentially understood by the difference of the contact distance (see Table~1).

For each reaction, let us define the fissibility (effective fission probability) as follows:
\begin{equation} \label{order} 
{\mathcal O} = \sum \sigma_{R_0} P_{fis,R_0},  \end{equation}
where $\sigma_{R_0}$ is the normalized cross-section satisfying $\sum_{R_0 \le R_C} \sigma_{R_0} = 1$ ($R_C$ is the contact distance), and $P_{fis,R_0} = 1, 0$ is the probability of fission; $P_{fis,R_0} = 1$ when the fission is calculated for the given initial distance $R_0$. 
Therefore $\sigma_{R_0}$ is the cross-section for realizing the initial state of TDDFT calculations with the initial distance $R_0$.
This parameter ($0 \le {\mathcal O} \le 1$), which is defined for each reaction (Eq.~\eqref{fiseq}) with fixed $^{\mathcal A}{\mathcal Z}$, $^{{\mathcal A}_1}{\mathcal Z}_1$ and $^{{\mathcal A}_2}{\mathcal Z}_2$, is the microscopic correspondence of the fissility $x$.
${\mathcal O}$ is equal to 0 if fission never appears, and to 1 if fission necessarily appears.
It is notable that, as is seen in the symmetric cases shown in Fig.~2, the fissibility can be sometimes equal to zero using this fissibility (cf. fissility $x$).
In order to clarify the difference depending on the mass asymmetry, let us introduce a simplified fissibility following Eq.~\eqref{order}.
\begin{equation} 
{\mathcal O}_{\rm simp} =  \max(P_{fis,R_0}).   
\end{equation} 
This parameter (${\mathcal O}_{\rm simp} =0,~1$) identifies if fission possibly appears (${\mathcal O}_{\rm simp} =1$) due to the collective dynamics (essentially the competition between the surface and Coulomb effects) or not (${\mathcal O}_{\rm simp} =0$).
Figure 3 shows the behaviour of the simplified fissibility within the performed calculations, where additional calculations 
with $R_0 =0,~2,~4,~6,~8$ fm
are made for three mass asymmetric cases in order to determine the value of ${\mathcal O}_{\rm simp}$.
Such a mass dependence cannot be described by the fissility $x$.
The investigation of the intermediate mass-asymmetry region ($0.16 < \min
({{\mathcal A}_1}/{{\mathcal A}_2},~{{\mathcal A}_2}/{{\mathcal A}_1}) < 1$)
will be studied in the future.

\section{Summary}  \label{sec3}
Compound nucleus fission is a complicated scenario in which the collective and single particle degrees of freedom, thermal effects 
(e.g., cooling effect leading to the localization), the collectivity due to the pairing, and other effects are activated on a 
case-by-case basis.
Among these factors, much attention has been paid to the collectivity in this paper.
A procedure of calculating the fissibility of compound nuclei has been presented.
The present method allows us to have a microscopic treatment for the fissibility for both symmetric and asymmetric events.
In particular, using this method, the shell effect, the deformation of the
fission fragments and the fermionic property (anti-symmetrization of many-body wave function) are introduced without having any additional treatments.

For testing, the method has been applied to analyse the fissibility of the compound nucleus $^{296}{\rm Lv}$.
The difference between symmetric and very asymmetric fissions has been clarified.  
In fact it turned out that very mass-asymmetric fissions cannot be explained by
the collective dynamics, although the competition between surface and Coulomb effects can be fully taken into account. 

\section*{Acknowledgements}
This work was supported by the Helmholtz alliance HA216/EMMI. The authors
thank Prof. J. A. Maruhn for reading this manuscript carefully.

\end{document}